\documentclass[12pt]{article}
\newcommand\mysection{\setcounter{equation}{0}\section}

\def\beq{\begin{equation}}
\def\eeq{\end{equation}}
\newcommand{\be}{\begin{equation}}
\newcommand{\ee}{\end{equation}}
\def\bea{\begin{eqnarray}}
\def\eea{\end{eqnarray}}

\def\1{{\rm 1 \kern -.10cm I \kern .14cm}} \def\R{{\rm R \kern -.28cm I
\kern .19cm}}

\def\be{\begin{equation}}
\def\ee{\end{equation}}
\newcommand\brr{\begin{eqnarray}}
\newcommand\err{\end{eqnarray}}

\begin{document}


\begin{titlepage}


\setcounter{page}{0}

\begin{flushright}
hep-th/0011223 \\
November 2000
\end{flushright}

\vspace{5 mm}
\begin{center}
{\large {\bf Some Issues in Noncommutative Solitons as $D$-branes}}

\vspace{20 mm}

{\large S. P. de Alwis, A. T. Flournoy\footnote{e-mail: 
dealwis@pizero.colorado.edu,flournoy@pizero.colorado.edu}}\\
{\em Department of Physics, Box 390,
University of Colorado, Boulder, CO 80309.}\\
\vspace{5 mm}
\end{center}

\vspace{10 mm}

\centerline{{\bf{Abstract}}}
We investigate and interpret a large class of  soliton solutions
found in noncommutative tachyon condensation.
These constructions make extensive use 
of the idea that $Dp$-branes may be built out of lower dimensional branes.  Finally we comment on
a recently proposed solution generating technique.
\end{titlepage}
\newpage
\renewcommand{\thefootnote}{\arabic{footnote}}
\setcounter{footnote}{0}


\mysection{Introduction}
\label{sec:introduction}

Sen has conjectured that the tachyonic vacuum in open bosonic string theory on a
D-brane describes the closed string vacuum without D-branes, and that various
soliton solutions in this theory describe D-branes of lower 
dimension\cite{sen1}.  Recent developments in noncommutative geometry
\cite{cds,dh,schomerus,seibergwitten} and in particular \cite{gms1} have
been tied in with this conclusion, resulting in a suprisingly solvable model
of D-brane annihilation.  The first steps in this recent progress
were made in the limit of an infinite background $B$-field
\cite{dmr,harvey1,witten}. Subsequently a  background independent formulation of the
model was proposed \cite{seiberg} following the reformulation of the original argument in \cite{gms2,sochichiu}.  This incorporated the advantages of
the infinite $B$ limit, while also offering some insight into
the vaccum structure of the closed string and connections to M(atrix) theory.
Though much progress has been made, a number of unresolveded issues remain.
These include various vacuum degeneracies,  inexplicable solitonic
configurations, the missing magnetic $21$-brane soliton, and unwanted
fluctuations on ``good'' solitonic solutions.
Speculations on the resolution of some of these issues have been put forward
\cite{hkl,sen2,sen3}.  In addition, a technique for generating solutions to
the equations of motion from known solutions has been proposed \cite{harvey3}.

In this paper we wish to expand on the idea \cite{seiberg} that these issues may be resolved
by identifying the underlying degrees of freedom as $D0$-branes or
$D$-instantons.  Though this is obviously closely related to a M(atrix) theory
\cite{matrix} interpretation, we will be working with the unstable branes of open bosonic
string theory.  Unlike the BPS $D0$-brane partons of M(atrix)
theory, our fundamental constituents of all $Dp$-branes will themselves be the
unstable $D0$-branes.  The instabilities of all of these $Dp$-branes are
reflected in the presence of a tachyonic open string state.

The paper is organized as follows.  In section
\ref{sec:theeffectivelagrangian} we write down the effective Lagrangian
governing the lowest modes in open string field theory.  Discussion then turns
in section \ref{sec:tensionsofsolitons} to the issue of soliton tensions and
how they essentially arise with the appearance of constituent
descriptions of $Dp$-branes.  Section \ref{sec:equationsofmotion} presents the
equations of motion arising from the effective Lagrangian and proposes a
simple set of conditions for identifying solutions.  The main discussion in
section \ref{sec:solutions} identifies both standard solutions to the
equations of motion and ``spurious'' ones, and motivates identifications of each
with underlying string/brane configurations.  Section
\ref{sec:alternativeaction} considers solutions arising from an alternative form of the
effective Lagrangian that has been proposed by several authors \cite{garousi,
  brwep, kluson}.  And the last
section discusses the recently proposed solution generating technique \cite{harvey3}.


\mysection{The Effective Lagrangian}
\label{sec:theeffectivelagrangian}

We begin with an expression for the effective Lagrangian on a single
Dp-brane in bosonic string theory after having integrated out all modes except the
tachyon and massless U(1) gauge field 
\be\label{} L(t)  = {1\over{(2\pi)^{{p-1}\over2}}}{1\over{g_s}} {\int
d^px [V(T)\sqrt{\det(g_{\mu \nu} + F_{\mu \nu})} -
f(T)\partial_{\mu}T\partial^{\mu}T\sqrt{detg_{\mu \nu}} + ...]}.
\label{eq:intlagrangian}\ee
We have set $2\pi \alpha' = 1$ and normalized $V(T)$ such that $V(T_{max})=1$.
With this normalization the coefficient in front of the action is exactly the
$D25$-brane tension.  The closed string background is defined by the closed string
metric $g_{\mu \nu}$, the closed string coupling $g_s$, and the constant background $B$-field.
To write the effective Lagrangian in the presence of a background $B$-field, we
need only begin with the effective Lagrangian without a $B$-field
(\ref{eq:intlagrangian}), replace the closed string parameters $g_{\mu \nu}$ and $g_s$ by
the corresponding open string parameters $G_{\mu \nu}$ and $G_s^2$, and further replace
all products of fields by the associative
$\star$-product of fields defined in terms of the noncommutativity parameter
$\Theta^{\mu \nu}$ \cite{seibergwitten}.  Whereas in (\ref{eq:intlagrangian}) the tachyon is neutral
under the $U(1)$ (as evidenced by the presence of ordinary derivatives in the
tachyon kinetic terms), the noncommutativity induces a nonzero coupling of the
tachyon to the gauge field.  Hence the ordinary derivatives of
(\ref{eq:intlagrangian}) should be replaced by gauge covariant ones.
An additional freedom in writing the effective Lagrangian arises from the choice of world-sheet
regularization which may be represented by a two form
$\Phi_{\mu \nu}$.  Putting these together we have
\be\label{} L(t)  = {1\over{(2\pi)^{{p-1}\over2}}}{1\over{G_s^2}} {\int
d^px [V(T)\sqrt{\det(G_{\mu \nu} +  F_{\mu \nu}^{NC} + \Phi_{\mu \nu})} -
f(T)D_{\mu}TD^{\mu}T\sqrt{detG_{\mu \nu}} + ...]_\star}.
\label{eq:intlagrangianB}\ee 

Upon choosing a particular $\Phi_{\mu \nu}$, we may then
express $G_{\mu \nu}$, $G_s$, and $\Theta^{\mu \nu}$ in terms of $g_{\mu \nu}$, $g_s$, $B_{\mu \nu}$,
 and $\Phi_{\mu \nu}$ using the relations
\begin{eqnarray} 
         ({1\over {G + \Phi}})^{\mu\nu} + \Theta^{\mu \nu}  &=&  ({1\over {g +B}})^{\mu \nu}  \\
         G_s^2  &=&  g_s \sqrt{{det(G +B)_{\mu \nu}}\over{det(g + B)_{\mu \nu}}} \nonumber
\end{eqnarray}\label{eq:phiequations}
Earlier work on the
subject\cite{harvey1,gms2} used the gauge $\Phi_{\mu \nu} = 0$.  We instead follow \cite{seiberg}
and take $\Phi_{\mu \nu} = -B_{\mu \nu}$ which leads to the exact expressions
\begin{eqnarray} 
         \Phi_{\mu \nu}  &=&  -B_{\mu \nu}    \nonumber \\ 
         \Theta^{\mu \nu}  &=&  ({1\over B})^{\mu \nu}  \\
         G_{\mu \nu}  &=&  -(B{1\over g} B)_{\mu \nu}  \nonumber \\
         G_s^2  &=&  g_s \sqrt{{detB_{\mu \nu}}\over{detg_{\mu \nu}}} \nonumber
\end{eqnarray}\label{eq:parameters}
These relations are precisely those found in the $\alpha'B_{\mu \nu} \to \infty$ limit of the
description in terms of $\Phi_{\mu \nu} = 0$.  In order to obtain manifest background
independence, we use as our gauge degrees of freedom the $X$ fields
\be\label{} X^\mu = x^\mu + \Theta^{\mu \nu} A_{\nu}^{NC}
\label{eq:xvariables}\ee
where the coordinates $x^{\mu}$ satisfy $[x^{\mu},x^{\nu}] = \Theta^{\mu \nu}$.  The effective
Lagrangian for a Dp-brane is then given by   
\be\label{} L(t)  = {1\over{(2\pi)^{{p-1}\over2}}}{1\over{g_s}} {\int
{d^px\over{\sqrt{det\Theta^{\mu \nu}}}} [V(T)\sqrt{\det(\delta_\mu^\nu +
  g_{\mu \lambda}[X^{\lambda},X^{\nu}])} -
f(T)g_{\mu \nu}[X^{\mu},T][X^{\nu},T] + ...]}.
\label{eq:intlagrangianX}\ee


\mysection{Tensions of Solitons}
\label{sec:tensionsofsolitons}

Utilizing ideas from \cite{gms1} the authors of \cite{dmr,harvey1} constructed
noncommutative solitons on the worldvolume of a bosonic $D25$-brane and
the work of Harvey et al \cite{harvey1} in fact identified them with $Dp$-branes of higher even codimension.  Part of the
evidence for this identification consisted of showing that the lower dimension
solitons exhibited tensions in agreement with the results for $Dp$-branes
obtained by T-duality.   Obtaining the correct tension though strongly suggestive, does not by itself constitute a complete demonstation that the solitionic configurations are the D-branes. This is because the reason that the  tension  comes out right has to do with the formula that gives the correspondence between functions of noncommuting coordinates and operators
on a Hilbert space.  In particular, for $\Theta^{\mu \nu}$ of rank $n$, we may group the
coordinates into $n$ noncommuting pairs with $26-2n$ leftover commuting
coordinates.  Functions of the $26$ coordinates can then be mapped to matrix-valued
functions of the $26-2n$ commuting coordinates.  The $\star$-product gets mapped to
the tensor product of operator multiplication with ordinary multiplication, and
most importantly the measure of integration over the noncommutative
coordinates gets mapped to a trace over the Hilbert space
\be\label{} {\int
{d^nx\over{\sqrt{det\Theta^{\mu \nu}}}}} \ldots \to (2\pi)^{n\over2}Tr \ldots
\label{eq:inttotrace}\ee

Considering the Lagrangian (\ref{eq:intlagrangianX}) for $p$-even.  We can
consider turning on a $B$-field in $p$ directions giving rise to a $\Theta^{\mu \nu}$ of rank
${p\over2}$.  Using the correspondence above (and restoring factors of
$2\pi\alpha'$) we would then have
\be\label{} L_p(t)  = {1\over{(2\pi)^p{(\alpha')^{{p+1}\over2}}}}{1\over{g_s}}
                    {\int {d^px\over{\sqrt{det\Theta^{\mu \nu}}}}} (\ldots)
                \longrightarrow {1\over{(2\pi)^p{(\alpha')^{{p+1}\over2}}}}{1\over{g_s}}
                    (2\pi)^{p\over2} (2\pi\alpha')^{p\over2} Tr (\ldots) 
                  = {1\over{{g_s\sqrt{\alpha'}}}} Tr (\ldots)
\label{eq:eventension}\ee
which may be identified with the Lagrangian for $N \to \infty$ $D0$-branes.

For $p$-odd, we can go to Euclidean space and consider turning on a $B$-field
in $p+1$ directions.  Using the operator correspondence, this leads to
\be\label{}    S_p  = {1\over{(2\pi)^p{(\alpha')^{{p+1}\over2}}}}{1\over{g_s}}
                    {\int {dtd^px\over{\sqrt{det\Theta^{\mu \nu}}}}} (\ldots)
                \longrightarrow {1\over{(2\pi)^p{(\alpha')^{{p+1}\over2}}}}{1\over{g_s}}
                    (2\pi)^{{p+1
}\over2} (2\pi\alpha')^{{p+1}\over2} Tr (\ldots) 
                  = {2\pi\over{g_s}} Tr (\ldots)
\label{eq:oddtension}\ee
which may be identified with the Euclidean action for $N \to \infty$
$D$-instantons.
It should be obvious by these identifications that obtaining the correct
tension for a soliton to be identified with a $Dp$-brane is built into the
formalism of describing functions of noncommuting coordinates by operators on
a Hilbert space.  The operator description is equivalent to a constituent
description.  The process of selecting a solitonic profile is merely to select a
subset of an infinitely extended collection of these constituent elements.
In light of this,  in order to properly  identify a configuration  its
fluctuation spectrum should also be considered. 
 

\mysection{Equations of Motion}
\label{sec:equationsofmotion}

To connect with issues raised in \cite{harvey1, gms2, seiberg} we consider the
effective action (\ref{eq:intlagrangianX}) for $p$=25.  For the explicit
construction of codimension-2n solitons it is only necessary to turn on a
background $B$-field in 2n directions\cite{harvey1}.  We will, however, be utilizing a maximal rank
$B$-field.  In doing so we will enable ourselves to deal with a more general
set of solutions than in \cite{harvey1}. To do so we work in Euclidean space. Turning on the
$B$-field in all 26 directions and using the operator correspondence outlined
in the previous section we have
\be\label{} S_{25}  ={ {2\pi}\over{g_s} }
{Tr [V(\hat T)\sqrt{\det(\delta_\mu^\nu + g_{\mu \lambda}[\hat X^\lambda,\hat X^\nu])} -
f(\hat T)g_{\mu \nu}[\hat X^\mu,\hat T][\hat X^\nu,\hat T] + ...]}.
\label{eq:trlagrangian}\ee
Defining
\be\label{} M_\mu^\nu \equiv \delta_\mu^\nu + g_{\mu \lambda}[\hat X^\lambda,\hat X^\nu]
\label{eq:defM}\ee
the tachyon equation of motion arising from (\ref{eq:trlagrangian}) is given by
\be\label{} V'(\hat T) {\sqrt{detM_\mu^\nu}} 
            - f'(\hat T) g_{\mu \nu} [\hat X^\mu,\hat T][\hat X^\nu,\hat T] 
             + g_{\mu \nu} [\hat X^\mu,[\hat X^\nu,\hat T]f(\hat T)] 
             + g_{\mu \nu} [\hat X^\nu,f(\hat T)[\hat X^\mu,\hat T] ] = 0
\label{eq:eomT}\ee
while the equation of motion from varying the $X$ field is
\be\label{} -{1\over2} [\hat X_\mu,(M^{-1} - (M^T)^{-1})^{\mu \nu} {\sqrt{detM_\mu^\nu}}
V(\hat T)] 
             + [\hat T,[\hat X^\nu,\hat T]f(\hat T)] 
             + [\hat T,f(\hat T)[\hat X^\nu,\hat T] ] = 0
\label{eq:eomX}\ee
To simplify matters we may consider a sufficient, but perhaps not necessary, 
set of conditions which will lead to solutions $\hat X_c$ and $\hat T_c$ of the 
equations above
\begin{eqnarray} 
         &a.& V'(\hat T) =  0 \nonumber  \\
         &b.& {[ \hat X^\mu , \hat T ]} =  0 \label{eq:conditions} \\
         &c.& {[ \hat X^\mu , [ \hat X^\nu , \hat X^\lambda ] ]} =  0 \nonumber
\end{eqnarray}
These conditions may not admit the complete set of solutions to the equations
of motion (\ref{eq:eomT},\ref{eq:eomX}), but we leave this additional
complication for future work.


\mysection{Solutions}
\label{sec:solutions}

We now consider solutions to these equations.  The first three represent
configurations that have a definite interpretation in terms of standard string/brane
configurations \cite{dmr,harvey1,gms2,seiberg}.  The remaining solutions have less obvious interpretations,
and as such must either be accounted for in standard string/brane configurations or
somehow excluded in this context.  One should keep in mind that the operator correspondence maps
the gauge covariant derivative in a particular direction to a commutator term
involving the $X$ field
\be\  D_\mu \hskip 0.25 cm \to {-i\Theta^{-1}_{\mu \nu} [\hat X^\nu,\hskip 0.7 cm ]},
\label{eq:derivative}\ee
so that if we have a background solution $\hat X^\mu = \lambda \hat I$, then propagating fluctuations in the
$x^\mu$ direction are forbidden.

\be \hat T_c = T_{max} \hat I  \hskip 1 cm  \hat X_c^\mu = \hat x^\mu  \hskip 1
cm \mbox{where} \hskip 0.75 cm  [\hat x^\mu,\hat x^\nu] = \Theta^{\mu \nu}\hat I. 
\label{eq:solution1}\ee 
This solution represents a uniform open string tachyon field on the 
worldvolume of an unstable $D25$-brane.  Fluctuations about this background
form a noncommutative $U(1)$ gauge theory with $26$-dimensional tachyons
transforming in the adjoint.  As a symmetry among the operators, the
noncommutative $U(1)$ is realized as a ${\displaystyle \bigotimes_{i=1}^{13}}U(N_i \to \infty)$.
The tension for this configuration may be identified by inserting the
background field configurations into the action;
\be\ S_{background} = {1\over{g_s}} {1\over{{(2\pi)}^{{25-1}\over2}}} \int {{d^{26}x}\over{\sqrt{det
    \Theta}}}  V(T_{max}) \sqrt{det(\delta_\mu^\nu + g_{\mu\lambda}\Theta ^{\lambda\nu})}. 
\label{eq:Sone}\ee
For $V(T_{max}) = 1$, we identify  the coefficient of the integral over the
$26$-dimensional worldvolume as the tension of the $D25$-brane.

\be \hskip 0.5 cm \hat T_c = T_{min} \hat I \hskip 1 cm  \hat X_c^\mu =  \lambda \hat I  
\label{eq:solution2}\ee
By Sen's conjecture, this uniform solution represents the stable closed string
vacuum in the absence of $D$-branes.  There are no propagating open string tachyon
fluctuations in this background.  The action with this background becomes 
\be\ S_{background} = {1\over{g_s}} {1\over{{(2\pi)}^{{25-1}\over2}}} \int {{d^{26}x}\over{\sqrt{det
    \Theta}}}  V(T_{min}) \sqrt{det(\delta_\mu^\nu)} 
\label{eq:Stwo}\ee
which vanishes according to Sen's conjecture, i.e. $V(T_{min}) = 0$.

\be \hskip 0.5 cm \hat T_c = T_{max} \hat P_n + T_{min} (\hat I - \hat P_n)  \hskip 1 cm 
  \hat X_c^i =  \hat x^i \hat P_n  \hskip 1 cm i=0,...,p  \hskip 1 cm  \hat X_c^m =
  \lambda \hat I \hskip 1 cm m=p+1,...,25 
\label{eq:solution3}\ee
The $\hat P_n$ in this expression are projection operators onto the Hilbert
space generated by the ``transverse'' noncommuting coordinates $x^m$.
The tachyon profile expressed in terms of $\hat P_n$ enjoys the
useful property that
\be\  V(\hat T_c) = V(T_{max}) \hat P_n + V(T_{min}) (\hat I - \hat P_n)
\label{eq:projector}\ee
These noncommutative solitons have finite extent in the $x^m$ directions and
infinite extent in the $x^i$ directions.  Such backgrounds interpolate in the
transverse directions $x^m$ between the tachyonic vaccum in the core of the
soliton and the closed string vacuum outside of the soliton.  Convincing
evidence has been put forward to identify these configurations with lower
dimensional unstable $Dp$-branes\cite{dmr,harvey1,gms2}.  To clarify this picture
we may consider a block diagonal noncommutativity matrix
\be\  \Theta = {\displaystyle \bigoplus_{k=1}^{13}}
      \left( \begin{array}{cc}
      o & \theta_k  \\
      -\theta_k & o  \end{array} \right) 
\label{eq:blocktheta}\ee
where
\be\  {[x^{2k},x^{2k + 1}]} = i\theta_k. 
\label{eq:thetapairs}\ee
Rank $n_k$ projection operators $P_{n_k}$ can be constructed on the Hilbert
spaces $H_k$ formed from each noncommuting pair of coordinates, so that a
general projection operator takes the form
\be\  P_n = P^{(1)}_{n_1} \otimes  P^{(2)}_{n_2} \otimes \ldots \otimes P^{({25 - p\over2})}_{n_{25 - p\over2}} 
\label{eq:projectors}\ee
for $p$ odd.  A soliton of this form corresponds to the chain of decays
\be\  1 D25 \to n_1 D23  \to n_1n_2 D21 \to \ldots \to {\displaystyle \prod_{k=1}^{25 - p\over2}} n_k Dp
\label{eq:branedecays}\ee
To concretely identify this configuration with $n$ coincident $Dp$-branes, we insert the solution
back into (\ref{eq:trlagrangian}) and use the correspondence
(\ref{eq:inttotrace}) on $p+1$ of the coordinates to obtain
\be\ S_{background} = {1\over{g_s}} {1\over{{(2\pi)}^{{p-1}\over2}}}
  Tr[\hat P_n V(T_{max}) + (\hat I - \hat P_n) V(T_{min})]
  \int {{d^{p+1}x}\over{\sqrt{det\Theta'}}} \sqrt{det(\delta_i^j + g_{ik}\Theta'^{kj})} 
\label{eq:Sthree}\ee
where
\be\  \Theta' = {\displaystyle \bigoplus_{k=1}^{p+1\over2}}
      \left( \begin{array}{cc}
      o & \theta_k  \\
      -\theta_k & o  \end{array} \right) 
\label{eq:blockthetaprime}\ee
If we use $V(T_{max})=1$ and Sen's conjecture,
i.e.$V(T_{min}) = 0$, then (\ref{eq:Sthree}) becomes
\be\ S_{background} = {n\over{g_s}} {1\over{{(2\pi)}^{{p-1}\over2}}}
  \int {{d^{p+1}x}\over{\sqrt{det\Theta'}}} \sqrt{det(\delta_i^j + g_{ik}\Theta'^{kj})} 
\label{eq:Sthreeprime}\ee 
which if compared to (\ref{eq:Sone}) is easily identified as the action for
$n$ $Dp$-branes. 

It should be noted that one can take the  limit $\theta_k\to 0$ in both (\ref{eq:Sone}) and (\ref{eq:Sthreeprime}) after redefining the coordinates  $x{2k,2k+1} \to {x^{2k,2k+1}\over{\sqrt{\theta_k}}}$ thus recoveing the usual form
of the D-brane action in the absence of guage fields or $B$ fields. Thus our procedure using the background independent formalism of \cite{seiberg} avoids the ambiguities associated with taking a large $B$ limit.

The solutions above admit simple and elegant interpretations in terms of
coincident unstable $Dp$-branes and the closed string vacuum.      However, other nonsingular solutions
to (\ref{eq:conditions}) exist, and thus solve the equations of motion arising
from (\ref{eq:trlagrangian}).  Since the action \ref{eq:intlagrangian} is expected to be an approximation  to the complete string field theory in the limit where the dervatives of the gauge fields and B fields are small, then these solutions must also be
accounted for as configurations of perturbative and nonperturbative states in
string theory.  The known perturbative and nonperturbative states of bosonic string
theory include the fundamental string and its magnetic dual, as well as
unstable $Dp$-branes for $p = -1,\ldots,25$.  However, the string field theory
action, if complete, could not only predict these states, but any possible
configuration of these states consistent with the background in which the
string field theory is formulated.  Certainly a number of the smooth solutions
to the full action will arise from its low energy effective form.  It is via
these nontrivial configurations
that we will interpret the additional solutions.  In our case this background is the
worldvolume of an unstable $D25$-brane with a constant maximal $B$-field.

A set of additional solutions to (\ref{eq:conditions}) was first pointed out
in \cite{gms2}.  These involve allowing different projection operators to
define the transverse profiles of the tachyon and $X$ fields.   To
systematically cover this set of solutions we will consider
descending from the $D25$-brane configuration by replacing the identity
operator by projectors where appropriate.  We first investigate the effect
of nontrivial projection operators for the tachyon while maintaining trivial
forms for $\hat X^L,\hat X^T$.  We then study the effects on  $\hat X^L$.
These results may be combined for configurations with nontrivial projection
operators for both $\hat T$ and $\hat X^L$. 

For now we consider operators
projecting onto the subspace generated by a single pair of coordinates which
we will refer to as simply $\hat x^T$, since these will in some sense be
interpreted as directions transverse to the resulting system.  The remaining
coordinates we refer to as $\hat x^L$ since these will be roughly longitudinal
to the system.  We consider functions of the coordinates $\vartheta(x^{\mu})$ which may be represented by
direct product operators $\hat \vartheta = \hat \vartheta_T \otimes \hat \vartheta_L$.  So we have for
the identity on the entire Hilbert space $H = H_L \otimes H_T$ an expression $\hat
I_{L,T} = \hat I_{L} \otimes \hat I_{T}$, and for a rank $n$ projection operator 
$\hat P_{n} = \hat I_{L} \otimes \hat P_{nT}$.

Consider a set of operators split into longitudinal and transverse parts.  The set of
solutions takes the form
\begin{eqnarray} 
         \hat T_c &=& T_{max}\hat I_L \otimes \hat P_{n1T} + T_{min}\hat I_L \otimes (
         \hat I_T - \hat P_{n1T})                \nonumber \\
         \hat X^{L}_c&=& \hat x^L \otimes \hat P_{n2T}   \label{eq:splitsolutions}          
\end{eqnarray}
The form of the transverse $X$ field determines two branches in the space of
solutions
\begin{eqnarray}
         1.\hskip 0.4 cm \hat X^{T}_c &=& \hat I_L \otimes \hat x^T \hskip 0.4 cm\mbox{with} \hskip 0.4 cm 
         \hat P_{n1T},\hat P_{n2T}\hskip 0.2 cm =\hskip 0.2 cm \hat 0^T\hskip 0.2 cm\mbox{or}\hskip 0.2 cm
         \hat I_T        \nonumber \\
         2.\hskip 0.4 cm \hat X^{T}_c &=& \lambda \hat I_L \otimes \hat I_T \hskip 0.4 cm\mbox{with
         }\hskip 0.4 cm {[\hat P_{n1T},\hat P_{n2T}]}\hskip 0.2 cm =\hskip 0.2 cm 0. \label{eq:twobranches}
\end{eqnarray}

We begin by descending the tachyon profile from $\hat I_{L,T} \to  \hat I_L \otimes
\hat P_{nT} \to \hat I_L \otimes \hat 0_T = \hat 0_{L,T}$ while keeping in mind
(\ref{eq:twobranches}) and holding $\hat X^{L}$ fixed
\be\label{} 
\begin{array}{cccc}
      a)
   &  \hat T_{c}  =  T_{max} \hat I_{L,T} 
   &  \hskip 1 cm \hat X^{L}_{c} = \hat x^L \otimes\hat I_{T}
   &  \hskip 1 cm \hat X^{T}_{c} = \hat I_{L} \otimes \hat x^T \\  
   &   \downarrow & \hskip 1 cm\downarrow & \hskip 1 cm\downarrow \\
      b)
   &  \hat T_{c}  =  T_{max} \hat I_{L,T} 
   &  \hskip 1 cm \hat X^{L}_{c} = \hat x^L \otimes\hat I_{T}
   &  \hskip 1 cm \hat X^{T}_{c} = \lambda \hat I_{L} \otimes \hat I_T  \\
   &   \downarrow & \hskip 1 cm\downarrow & \hskip 1 cm\downarrow \\
      c)  
   &  \hat T_{c}  = T_{max}\hat I_L \otimes \hat P_{nT}
    + T_{min} \hat I_L \otimes(\hat I_T  - \hat  P_{nT}) 
   &  \hskip 1 cm \hat X^{L}_{c} = \hat x^L\otimes\hat I_{T}
   &  \hskip 1 cm \hat X^{T}_{c} = \lambda \hat I_{L} \otimes \hat I_T \\
   &   \downarrow & \hskip 1 cm\downarrow & \hskip 1 cm\downarrow \\
      d)
   &  \hat T_{c}  = T_{min} \hat I_{L,T}  
   &  \hskip 1 cm \hat X^{L}_{c} = \hat x^L \otimes\hat I_{T}
   &  \hskip 1 cm \hat X^{T}_{c} = \lambda \hat I_{L} \otimes \hat I_T \\
   &   \\
      e)
   &  \hat T_{c}  = T_{min} \hat I_{L,T}  
   &  \hskip 1 cm \hat X^{L}_{c} = \hat x^L \otimes\hat I_{T}
   &  \hskip 1 cm \hat X^{T}_{c} = \hat I_{L} \otimes \hat x^T 
\end{array}  
\label{eq:tachyondescent}\ee
One should keep in mind that the tachyon profile will always lead to a simple
tension expression which can then be used to identify the $Dp$-brane
present.  The $X$ field configuration on the other hand governs the
propagation of various fluctuations, and so should give us information on how
the constituent $Dp$-branes are assembled.

The process $a \to b$ represents a transition from a space filling $D25$-brane
into a space filling stack in the $x^T$ directions of an infinite number of $D23$-branes with
worldvolume extension in the $x^L$ directions, and with open strings confined to
each constituent brane.  That open string modes cannot propagate in $x^T$ is a
consequence of the vanishing of the covariant derivative in the transverse
directions  
\be\label{}
D_T \hskip 0.25 cm \to {-i \Theta^{-1}_{\mu \nu}[\lambda \hat I_L \otimes \hat I_L,\hskip 0.7 cm
  ]} = 0
\label{eq:zerotransversederivative}\ee
In essence, propagation in the $x^T$ directions is eliminated by not allowing
open strings to migrate from one $D23$-brane to the next.  Propagation of
fluctuations along the $D23$-branes is of course still allowed.

The transition $b \to c$ represents the decay of $\infty - n$ of the $D23$-branes into the closed
string vacuum.  In contrast to the standard $D23$-brane solution which
implements the same projection operator for the tachyon and longitudinal $X$
fields, here we have allowed a nontrivial 
projection operator for the tachyon alone.  However, the resulting
configuration is physically indistiguishable from the standard $D23$-brane
solution (\ref{eq:solution3}). 
This is a simple consequence of the factors of $V(T)$ and $f(T)$ in front of the
Born-Infeld and tachyon kinetic terms respectively. These give rise to an overall factor of
the tachyon projection operator which acts on the $X$ fields,
effectively giving $\hat X^L$ the same projector form as $\hat T$.  The action
evaluated for this background is precisely (\ref{eq:Sthree}).  In this case (with the projector
profile) we can not reinstate propagation in the $x^T$ directions in light of
(\ref{eq:twobranches}).  This is a reflection of the simple
fact that a $D25$-brane cannot be constructed out of a finite number of
$D23$-branes. 

The transition $c \to d$ seems to have no definite interpretation unless the
assumption $f(T_{min} = 0$ is made.  With this assumption configuration $d$ is
physically identical to the closed string vacuum (\ref{eq:solution2}).
Without this assumption the tachyon
profile will yield a vanishing tension, yet propagating fluctuations are
allowed along the $x^L$ directions.  One should note however that with this
choice for the tachyon field, the overall coefficient $V(T_{min})$ of the
Born-Infeld contribution to the action vanishes.  We expect that our
computations, based on a well defined action, will run into trouble in this scenario.

The transition $a \to e$ shares the complication of $a \to b$ and will need
further investigation of the action to be interpreted or excluded.  Sen has
argued \cite{sen2} that configurations $d$ and $e$ are actually
equivalent descriptions of the closed string vacuum. 

We may now consider the results of descending the longitudinal $X$ field
$\hat X^L$, this time holding fixed the tachyon while maintaining 
(\ref{eq:twobranches}).   

\be\label{} 
\begin{array}{cccc}
      a)
   &  \hat T_{c}  =  T_{max} \hat I_{L,T} 
   &  \hskip 1 cm \hat X^{L}_{c} = \hat x^L \otimes\hat I_{T}
   &  \hskip 1 cm \hat X^{T}_{c} = \hat I_{L} \otimes \hat x^T  \\
   &   \downarrow & \hskip 1 cm\downarrow & \hskip 1 cm\downarrow \\
      b)
   &  \hat T_{c}  =  T_{max} \hat I_{L,T} 
   &  \hskip 1 cm \hat X^{L}_{c} = \hat x^L \otimes\hat I_{T}
   &  \hskip 1 cm \hat X^{T}_{c} = \lambda \hat I_{L} \otimes \hat I_T  \\
   &   \downarrow & \hskip 1 cm\downarrow & \hskip 1 cm\downarrow \\
      c)  
   &  \hat T_{c}  = T_{max}\hat I_{L,T}
   &  \hskip 1 cm \hat X^{L}_{c} = \hat x^L\otimes\hat P_{nT}
   &  \hskip 1 cm \hat X^{T}_{c} = \lambda \hat I_{L} \otimes \hat I_T \\
   &   \downarrow & \hskip 1 cm\downarrow & \hskip 1 cm\downarrow \\
      d)
   &  \hat T_{c}  = T_{max} \hat I_{L,T}  
   &  \hskip 1 cm \hat X^{L}_{c} = \hat I_L\otimes\hat 0^T \cong \lambda \hat I_{L} \otimes \hat I_T
   &  \hskip 1 cm \hat X^{T}_{c} = \lambda \hat I_{L} \otimes \hat I_T \\ 
   &  \\
      e)
   &  \hat T_{c}  = T_{max} \hat I_{L,T}  
   &  \hskip 1 cm \hat X^{L}_{c} = \hat I_L\otimes\hat 0^T \cong \lambda \hat I_{L} \otimes \hat I_T
   &  \hskip 1 cm \hat X^{T}_{c} = \hat I_{L} \otimes \hat x^T 
\end{array}  
\label{eq:longgaugedescent}\ee
We identify in $c,d$ that taking the longitudinal $X$ field projection
operator to have rank $0$ is effectively equivalent to setting $\hat X^L \propto \hat I$.
 
The process $a \to b$ is identical to the case discussed for
(\ref{eq:tachyondescent}), that is the transition from a $D25$-brane into an infinite
number of $D23$-branes with open string ends restricted to single $D23$-branes.

The transition $b \to c$ is rather interesting.  The tension is that of the
space filling $D25$-brane.  The propagation of fluctuations in the transverse
directions $x^T$ is eliminated as in configuration $b$.  However, now the
propagation of fluctuations in the longitudinal directions $x^L$ is restricted
to a region localized around the origin in $x^T$ as described by $\hat
P_{nT}$.  The interpretation of this is as follows: each of $\infty - n$ of the 
infinite collection of $D23$-branes in configuration b undergoes a transition
to an infinite collection of $D$-instantons with the ends of any open string
confined to a single $D$-instanton.  This confinement eliminates propagating
fluctuations in the longitudinal directions $x^L$ outside of a region
localized in $x^T$.  The remaining $n D23$-branes are stacked
as before about the origin in $x^T$ and admit propagating fluctuations along
$x^L$. 

The transition $c \to d$ corresponds to a transition of the remaining
$D23$-branes in configuration $c$ to infinite collections of $D$-instantons.
Since the tachyon profile has not changed, we expect an energy density
identical to that of a $D25$-brane.  In fact the action evaluated with this
background field configuration is essentially that of (\ref{eq:oddtension}), i.e.
\be\ S_{background} = {{2\pi}\over{g_s}} Tr \hat I_{L,T} =  {{2\pi}\over{g_s}}
(N \to \infty)^{13}
\label{eq:Sfour}\ee
Since the ends of open strings are confined to points in
space-time, there are of course no propagating open string modes.

The transition $a \to e$ is actually very closely related to the $a \to b$
transition.  In this case it may seem that the transition of the space filling $D25$-brane is to an
infinite collection of $D1$-branes.  However, in interpreting configuration
$c$ which is extended in the $x^L$ directions
as a stack of coincident $D23$-branes, we are assuming that time is an element of the
$x^L$ set of coordinates.  Thus for configuration $e$ we have a stack of
two-dimensional objects extended in purely spatial directions $x^T$ and exist
only at an instant in time.  These
extended instantons correspond to objects resulting from $T$-dualizing 
the time coordinate on a $D2$-brane.  Whether this operation is well defined
and the role played by the resulting objects is outside the scope of this
paper. For a discussion of such issues see \cite{hull} and references therein.

The two descent chains above may be used as the cornerstones for more
complicated projector combinations.  In any case one is led to a description
of the $D25$-brane in terms of some constituent $Dp$-branes which are either
decayed, transversely distributed, coincident, or some combination of these
which may differ for different directions.

A final set of solutions that may be considered are those which utilize
operators on the transverse subspace other than $\hat I, \hat P_n , \hat 0, \hat x^T$.  These must commute
with any other operator acting in this subspace in order to serve
as solutions to (\ref{eq:conditions}).  The operators above are distinguished
by possessing eigenvalues equal to either $0$ or $1$, for $\hat I, \hat P_n ,
\hat 0$ and values filling out the real line for $\hat x^T$.  Commuting operators
with more diverse spectra of eigenvalues certainly exist.  To this end we may
consider solutions of the form

\be \hskip 0.5 cm \hat T_c = T_{max}\hat I_L \otimes \hat P_{n1T} + T_{min}\hat I_L \otimes (
      \hat I_T - \hat P_{n1T})
      \hskip 1 cm  \hat X^{L}_c = \hat x^L \otimes \hat M  
      \hskip 1 cm  \hat X^T_c = 0
\label{eq:solution5}\ee

\noindent where

\be
     \hat M \equiv {\displaystyle \sum_{i=1}^{k}} \lambda_i \hat P^i_{n_i}
     \hskip 0.5 cm \mbox{for}
     \hskip 0.5 cm \hat P^i_{n_i} \hat P^j_{n_j} = \hat P^i_{n_i} \delta^{ij}
     \hskip 0.5 cm \mbox{and}
     \hskip 0.5 cm {[\hat M , \hat P_{n1T}]} = 0
\label{eq:solution5extra}\ee

\noindent We may simplify matters by assuming 
\be
     Tr \hat M = Tr \hat P_{n1T}
     \hskip 0.5 cm \hat M \hat P_{n1T} = \hat M.
\label{eq:solution5cond}\ee 
to avoid the complications discussed after equation (\ref{eq:splitsolutions}).

\noindent For all $\lambda_i$ distinct, this configuration maintains only a 
${\displaystyle \bigotimes_{i=1}^{k}}U(n_i)$ of the $U({\displaystyle 
\sum_{i=1}^{k}} n_i)$ present in the most degenerate case.  From the gauge
theory point of view this symmetry breaking corresponds to a separation of the
branes.  Interpreting the separation in eigenvalue space as a spacetime
distance {\`a} la M(atrix) theory \cite{matrix}, this
configuration can be identified with a collection of $k$ non-coincident  
stacks of $D23$-branes.


\mysection{Alternative action}
\label{sec:alternativeaction}

By demanding consistency with T-duality several authors \cite{garousi, brwep, kluson}
have obtained alternate forms of our starting point (\ref{eq:intlagrangian})
which differ by including the tachyon kinetic term under the square root 
\be\label{} L(t)  = {1\over{(2\pi)^{{p-1}\over2}}}{1\over{g_s}} {\int
d^px  V(T)\sqrt{\det(g_{\mu \nu} + F_{\mu \nu} + \partial_\mu T\partial_\nu T )} }.
\label{eq:intaltlagrangian}\ee   
An obvious advantage in considering actions of this form is the
automatic vanishing of the tachyon kinetic term for $T = T_{min}$.  We may
repeat the analysis above for this modified effective Lagrangian.   
Turning on a maximal rank $B$-field and using the operator correspondence we
obtain
\be\label{} S_{25}  ={ {2\pi}\over{g_s} }
{Tr V(\hat T)\sqrt{\det(\delta_\mu^\nu + g_{\mu \lambda}[\hat X^\lambda,\hat X^\nu]
 + g_{\mu \nu}[\hat X^\mu, \hat T][\hat X^\nu, \hat T])}}.
\label{eq:traltlagrangian}\ee
Define
\be\label{} W \equiv \delta_\mu^\nu + g_{\mu \lambda}[\hat X^\lambda,\hat
X^\nu] + g_{\mu \lambda}[\hat X^\lambda, \hat T][\hat X^\nu, \hat T].
\label{eq:defW}\ee
The tachyon equation of motion is now given by
\be\label{} V'(\hat T) {\sqrt{detW_\mu^\nu}} 
             + [\hat X_\mu,[\hat X_\nu,\hat T]]W^{-1 \mu \nu}{\sqrt{detW_\mu^\nu}}V(\hat T) 
             - [\hat X_\nu,W^{-1 \mu \nu}{\sqrt{detW_\mu^\nu}}V(\hat T)[\hat X_\mu,\hat T]] 
             = 0
\label{eq:eomaltT}\ee
while the equation of motion from varying the $X$ field is
\begin{eqnarray}
          [\hat X_\mu,(W^{-1} - (W^T)^{-1})^{\nu \mu} {\sqrt{detW_\mu^\nu}}V(\hat T)] 
             + [\hat T,[\hat X_\mu,\hat T]W^{-1 \mu \nu}{\sqrt{detW_\mu^\nu}}V(\hat
             T)] \nonumber \\
             + [\hat T,W^{-1 \nu \mu}{\sqrt{detW_\mu^\nu}}V(\hat T) [\hat X_\mu,\hat T] ] 
             = 0. \label{eq:eomaltX}
\end{eqnarray}
Again we may consider a set of sufficient conditions for a solution of these
equations
\begin{eqnarray} 
         V'(\hat T_c) &=&  0    \nonumber \\
         \label{}   
         {[ \hat X_c^\mu , [\hat X_c^\nu,\hat T_c]]} &=&  0 \label{eq:altconditions} \\
         {[ \hat X_c^\mu , [ \hat X_c^\nu , \hat X_c^\lambda ] ]} &=&  0.  \nonumber
\end{eqnarray}
These conditions admit all of the configurations discussed in the previous
section as solutions since $\hat X_c^\nu$ and $\hat T_c$ satisfying
\be\label{}   {[\hat X_c^\nu,\hat T_c]} =  0
\label{eq:oldcom}\ee
certainly describe a subset of the solutions of (\ref{eq:altconditions}).
However, the conditions (\ref{eq:altconditions}) admit now a larger set of
solutions including, for example, configurations satisfying
\be\label{}   {[\hat X_c^\nu,\hat T_c]} \propto  \hat I.
\label{eq:altcom}\ee
Such solutions are considerably more difficult to explicitly construct than those in
the preceding discussion.  However, owing to the advantage of the
automatically vanishing
tachyon kinetic term for $T = T_{min}$ it would be worthwile to investigate
these solutions further.


\mysection{The Shift Operation}
\label{sec:theshiftoperation}

Recently a technique was introduced to facilitate finding solutions to the 
equations of motion for noncommutative guage theories from known solutions by
acting with an ``almost'' gauge transformation \cite{harvey3}.  This method
was applied to vacuum solutions in open string field theory to obtain
solitonic field configurations which might then be interpreted as $Dp$-branes.
There are a few issues regarding this construction which we feel should be
discussed.
\begin{itemize}

\item The shift operation

Formulation of the solution generating technique began by observing that a
transformation obeying
\begin{eqnarray} 
           \hat U^{\dagger} \hat U = \hat I  \nonumber \\
          \label{}   
           \hat U \hat U^{\dagger}  = \hat P \label{eq:shifts} 
\end{eqnarray}
where $\hat P$ is a projection operator, when applied to the fields $\hat \vartheta^i$ in an 
equation of motion would result in new field configurations obeying the same
equation of motion.

\item Tensions and the tachyon
 
Solutions to (\ref{eq:shifts}) only exist for infinite dimensional $\hat U$.  The authors
of \cite{harvey3} construct an infinite dimensional representation with the
shift operators
\be\label{}  \hat S =  {\displaystyle  \sum_{k=0}^{\infty}}|k+1><k| 
\label{eq:shiftop}\ee  
which satisfy
\be\label{}  \hat S^{n \dagger}\hat S^n = I, \hskip 1 cm  \hat S^n\hat S^{n \dagger} = \hat I - \hat P_n
\label{eq:shiftopcond}\ee
where $\hat P_n$ are projection operators onto the first $n$ states.
The effect of $\hat U = \hat S^n$ on the matrix representation of a field $\hat \vartheta$ is a 
``southeast shift''.
The idea proposed in \cite{harvey3} is that by acting on the closed string
vacuum field configurations with $\hat U$ defined above one may generate
configurations corresponding to $Dp$-branes.  To see this in action, we will
look at the effect of $\hat U = \hat S^n$ on the vaccum tachyon field configuration
discussed in section \ref{sec:solutions}.  We will merely consider trying to build
a pair of $D23$-branes from the closed string vacuum, so that the corresponding
projection operator is nontrivial in the subspace $H_k$ generated by
$[\hat x^{24},\hat x^{25}] = i \theta\hat I$.  The tachyon vacuum configuration transforms as follows
\be\label{} \hat T_{vac} = 
\left( \begin{array}{ccccc}
       T_{min} & 0 & 0 & 0 & \ldots\\
       0 & T_{min} & 0 & 0 & \ldots\\
       0 & 0 & T_{min} & 0 & \ldots\\
       0 & 0 & 0 & T_{min} & \ldots\\
        \vdots & \vdots & \vdots & \vdots &\ddots  \end{array} \right) 
\longrightarrow \hat S^2\hat T_{vac}\hat {S}^{2 \dagger} = 
\left( \begin{array}{ccccc}
       0 & 0 & 0 & 0 & \ldots\\
       0 & 0 & 0 & 0 & \ldots\\
       0 & 0 & T_{min} & 0 & \ldots\\
       0 & 0 & 0 & T_{min} & \ldots\\
       \vdots & \vdots & \vdots & \vdots &\ddots  \end{array} \right)
\label{eq:shifttachyon}\ee
The resulting configuration may be identified with the tachyon configuration
corresponding to a pair of $D23$-branes
\be\label{} \hat T_{D23} = 
\left( \begin{array}{ccccc}
       T_{max} & 0 & 0 & 0 & \ldots\\
       0 & T_{max} & 0 & 0 & \ldots\\
       0 & 0 & T_{min} & 0 & \ldots\\
       0 & 0 & 0 & T_{min} & \ldots\\
        \vdots & \vdots & \vdots & \vdots &\ddots  \end{array} \right) 
\label{eq:D1tachyon}\ee
only if we have arranged that $T_{max} = 0$. For this mechanism to
work for a choice of $T_{max} \neq 0$, it would be necessary for the shift to
produce an upper left diagonal block $diag(T_{max} , \ldots , T_{max})$.  However, a ``shift'' operation
accommodating nonzero ``northwest'' elements can not be constructed, and so
this procedure exhibits a peculiar dependance on the value of what one might have expected 
to be an arbitrary choice.  If the choice
$T_{max} = 0$ is made, then one obtains the correct tension for the $D23$
pair in light of (\ref{eq:projector}).

The point is that an equation of motion of the form $F(\vartheta )=0$ will give rise to an equation $\hat UF(\vartheta )\hat U^{\dag}=0$ under the action of the shift transformation. But this is not the same as $F(\hat U\vartheta \hat U^{\dag})=0$ unless
$F(0)=0$ is true as well.

\item $X$ fields

Let us now investigate the result of the shift transformation on the gauge
field configurations corresponding to the closed string vacuum in our
formalism.  Applying $\hat U = \hat S^2$ to $\hat X^\mu_{vac}$ we have
\be\label{} \hat X^\mu_{vac} = \hat 0 
\longrightarrow \hat S^2\hat X^\mu_{vac}\hat {S}^{2 \dagger} = \hat 0 
\label{eq:shiftgauge}\ee
where $\hat 0$ represents the null matrix.  The result
above will pose a problem when we compare the
shifted $\hat X^\mu_{vac}$ to the expected $X$ field configuration for a pair of
$D23$-branes (see (\ref{eq:solution3}))
\be\label{} \hat X^i_{D23} = 
\left( \begin{array}{ccccc}
       \hat x^i & 0 & 0 & 0 & \ldots\\
       0 & \hat x^i & 0 & 0 & \ldots\\
       0 & 0 & 0 & 0 & \ldots\\
       0 & 0 & 0 & 0 & \ldots\\
        \vdots & \vdots & \vdots & \vdots &\ddots  \end{array} \right)  
\hskip 0.25 cm i = 0,\ldots,23 \hskip 1.5 cm
\hat X^m_{D23} = 
\left( \begin{array}{ccccc}
       0 & 0 & 0 & 0 & \ldots\\
       0 & 0 & 0 & 0 & \ldots\\
       0 & 0 & 0 & 0 & \ldots\\
       0 & 0 & 0 & 0 & \ldots\\
       \vdots & \vdots & \vdots & \vdots &\ddots  \end{array} \right)
\hskip 0.25 cm m = 24,25 
\label{eq:D1gauge}\ee
It appears that in order for this technique to work the shift operation
would have to distinguish between components of the $X$ field, and produce a
nonzero ``northwest'' block for the components along the brane.  In addition, the 
transformation would have to distinguish between the tachyon and $X$ fields
 and produce appropriate ``northwest'' blocks for each.  

This issue does not arise in \cite{harvey3}.  In that
work a choice is made for the closed
string vacuum configuration which effectively reverses the situation above,
that is
\be\label{} \hat X^\mu_{vac'} = \hat x^{\mu}. 
\label{eq:harveygaugevac}\ee
$X$ field configurations for the $D$-branes are identified as
\be\label{} \hat X^i_{D23'} =
\left( \begin{array}{ccccc}
       0 & 0 & 0 & 0 & \ldots\\
       0 & 0 & 0 & 0 & \ldots\\
       0 & 0 & \hat x^i & 0 & \ldots\\
       0 & 0 & 0 & \hat x^i & \ldots\\
        \vdots & \vdots & \vdots & \vdots &\ddots  \end{array} \right)  
\hskip 0.25 cm i = 0,1 \hskip 0.75 cm
\hat X^m_{D23'} =
\left( \begin{array}{ccccc}
       \hat x^m & 0 & 0 & 0 & \ldots\\
       0 & \hat x^m & 0 & 0 & \ldots\\
       0 & 0 & \hat x^m & 0 & \ldots\\
       0 & 0 & 0 & \hat x^m & \ldots\\
       \vdots & \vdots & \vdots & \vdots &\ddots  \end{array} \right)
\hskip 0.25 cm m = 2,\ldots,25 
\label{eq:harveyD1gauge}\ee
Again, the solution generating transformation seems to depend on the choice of
$0$ for the diagonal $X$ field terms representing $D$-branes.  This
identification arises directly from the initial choice
(\ref{eq:harveygaugevac}) for the vacuum $X$ field configurations.  In
\cite{harvey3} it is proposed that these configurations afford an extension of
the $\hat X^\mu = 0$ vacuum to appropriate configurations for arbitrary
noncommutativity $\theta$.  These configurations reproduce the correct
expressions for $D$-brane tensions, however as discussed in section
\ref{sec:tensionsofsolitons}, the evidence for identifying these as the 
correct configurations should take into account the spectrum of fluctuations
as well.  The configuration
(\ref{eq:harveygaugevac}) will admit a spectrum of fluctuations identifiable
with that of the closed string vacuum if one of two conditions hold.  Either
one works in the $\alpha'B \to \infty$ limit or one conjectures that the
coefficient function for the tachyon kinetic term in the action vanishes for
$\hat T = T_{min}\hat I$, i.e. $f(T_{min}) = 0$. The actions discussed in
section \ref{sec:alternativeaction} automatically enforce the latter of these.

\end{itemize}


\mysection{Conclusions}
\label{sec:conclusions}

The explicit construction of $Dp$-branes as lumps in open string field theory
and the emergence of the closed string vacuum provides a strong indication
that a string field formulation may provide a complete nonperturbative definition of
string theory.  Of course we have limited ourselves to the simple setting of
the bosonic string, but complications that must be dealt with in the bosonic
theory are sure to arise in the supersymmetric formulation as well.  

As a nonperturbative formulation, the string field action should predict all
of the known perturbative and nonperturbative states in string theory.  In
addition, we expect that all solutions of the string field equations of motion
should find some interpretation in terms of string/brane configurations.  We
have demonstrated that a large class of the explicit solutions constructed via
techniques from noncommutative geometry have such an interpretation.  A main
feature of these constructions involved viewing higher dimensional $Dp$-branes
to be composed of lower dimensional $Dp$-branes.  In particular,
infinite configurations of $D$-instantons allow one to account for the tension of
higher dimensional branes, while forbidding propagating fluctuations. Though
the general solutions become very complicated very quickly, we expect that
the simple ideas presented here can be used to construct any configuration
required.  

Extending the bosonic string field effective action to include modifications
consistent with $T$-duality seems to enlargen the space of solutions.
Explicit construction of these new solutions is difficult, and their
interpretations will certainly not be straight forward.  On the other hand an
advantage to this form of the action is that the tachyon kinetic term automatically
vanishes for $T = T_{min}$.  Future work may involve analyzing and
interpreting these new configurations.

The shift symmetry solution generating technique may provide some important
insight into noncommutative tachyon condensation. There is certainly an appeal to the generation of solutions from solutions via
a single well defined transformation.  The construction is reminiscent of $T$-duality  and it would be nice to have a better understanding of  the issues discussed in section 7.

\vskip 1cm 
{\bf Acknowledgements}
\vskip .5cm

This work is partially supported by the Department of
Energy contract No. DE-FG02-91-ER-40672. One of us wishes to thank
the Aspen center for physics for hospitality and Per Kraus for discussions in the early stages of this work.
The authors also wish to thank Finn Larsen for discussions on \cite{harvey3} and for useful comments on a preliminary version of this work.


{}

\end{document}